\documentclass[]{aastex631}

\submitjournal{\apj}
\accepted{January 30, 2023}

\shorttitle{Interstellar Contrails}
\shortauthors{Kitajima \& Inutsuka}
\graphicspath{{./}{figures/}}
\begin{document}

\title{An origin of narrow extended structure in the interstellar medium:\\
an interstellar contrail created by a fast-moving massive object}

\author[0000-0003-0738-0707]{Kanta Kitajima}
\altaffiliation{Email: {\tt kitajima.kanta.z7@s.mail.nagoya-u.ac.jp}}
\affiliation{Department of Physics, \\
Graduate School of Science, \\
Nagoya University, \\
Furo-cho, Chikusa-ku, Nagoya  464-8692, Japan}

\author[0000-0003-4366-6518]{Shu-ichiro Inutsuka}
\affiliation{Department of Physics, \\
Graduate School of Science, \\
Nagoya University,  \\
Furo-cho, Chikusa-ku, Nagoya  464-8692, Japan}

\begin{abstract}
We investigate the thermal condensation caused by
a massive object that passes through the interstellar medium with high velocity, and propose a mechanism for creating a filamentary gaseous object, or interstellar contrail. 
Our main result shows that a long interstellar contrail can form with a certain parameter; a compact object more massive than $10^4\ {\rm M_\odot}$ can make a filament whose length is larger than $100\ {\rm pc}$.
Observation of interstellar contrails may provide information on the number, masses, and velocities of fast-moving massive objects, and can be a new method for probing invisible gravitating sources such as intermediate-mass black holes.

\end{abstract}

\section{Introduction} \label{sec:intro}
A sizable fraction of interstellar gas is in approximate thermal equilibrium where heating and cooling are balanced \citep[e.g.,][]{Ferriere_1998}.
The stable phases are classified as the cold neutral medium (CNM) and the warm neutral medium (WNM)
\citep[e.g.,][]{Field_etal_1969,Cox_2005}.
When a shock wave propagates through such a stable gas, it is compressed by thermal condensation \citep[e.g.,][]{Field_1965,Falle_etal_2020}.
Supernovae are supposed to be the main cause of those shock waves, and their interactions with the interstellar medium (ISM) have been extensively studied
\citep[e.g.,][]{Koyama_Inutsuka_2000,Koyama_Inutsuka_2002,Audit_Hennebelle_2005,Inoue_Inutsuka_2008,Inoue_Inutsuka_2009,Kupilas_etal_2021}.

We propose that a massive compact object that moves fast creates a filamentary structure (interstellar contrail). 
This is because by gravitational focusing due to the massive compact object moving fast, a shock wave is driven and thermal condensation occurs. 
Although it has been suggested that trajectories of celestial objects create structure in interstellar space \citep[e.g.,][]{Wallin_etal_1996,Martinez-Medina_etal_2016,Li_Shi_2021}, quantitative analysis of the resultant thermal structure has not been studied. 
This paper analyzes the cloud formation quantitatively. 
Note that the structure analyzed in this paper is very different from the bow shock and cocoon generated by a supersonically moving object that has a surface \citep[e.g.,][]{Wilkin_1996}.

In the next section, we explain our model and the details of our analysis.
Results are presented in \S3.
In \S4, we discuss approximate formulates of an interstellar contrail and the possibility of observation.
\S5 presents a summary.
The symbols and their definitions that we use in this paper are compiled in Tab.\ref{table:data_type}.

\begin{table}[hbtp]
 \caption{Notation used in This Paper}
  \label{table:data_type}
  \centering
  \begin{tabular}{ll}
    \hline
    Symbol & Definition\\
    \hline \hline
    $A$ & Particle number accretion rate per unit length of filament\\
    $b$ & Impact parameter\\
    $\tilde{b}$ & Nondimensionalized $b$ with $r_{\rm HL}$\\
    $b_{\max}$ & Impact parameter corresponding to $L_{\rm pt}$\\
    $c_{\rm pre}$ & Speed of sound in front of the shock wave\\
    $d\dot{M}$ & Mass accretion rate per unit length\\
    $L$ & Filament length\\
    $L_{\rm pt}$ & Largest $z$ coordinate where the gas density reaches high enough to be observed by thermal condensation\\
    $L_{\rm pt}'$ & Largest $z$ coordinate where the gas density reaches high enough to be observed by thermal condensation \\
    & obtained without considering the condition for shock wave generation\\
    $L_{\rm sh}$ & Largest $z$ coordinate where a shock wave is generated\\
    $m_{\rm p}$ & Mass of proton\\
    $n_{\rm f}$ & Number density of gas in the filament\\
    $n_{\rm pre}$ & Number density of gas in front of the shock surface\\
    $n_\infty$ & Number density of gas in the ambient medium\\
    $P_{\rm crit}$ & Critical pressure of thermal condensation\\
    $P_{\rm eq}$ & Pressure of HI gas in thermal equilibrium\\
    $M$ & Mass of the gravitating object\\
    $\dot{M}$ & Mass accretion rate to the gravitating object\\
    $M_{\rm f}$ & Total mass of filament\\
    $\dot{M}_{\rm f}$ & Mass accretion rate on the filament\\
    $\tilde{r}$ & Radius of a fluid element in units of  $r_{\rm HL}$\\
    $r_{\rm HL}$ & Hoyle--Lyttleton accretion radius\\
    $R$ & Position of filament surface\\
    $\tilde{R}$ & Filament radius in units of  $r_{\rm HL}$\\
    $R_{\rm max}$ & Radius at the end of the filament\\
    $s$ & Position of the stagnation point\\
    $T_{\rm eq}$ & Gas temperature in thermal equilibrium\\
    $T_{\rm post-shock}$ & Gas temperature in the post shock region\\
    $v_{\rm FC}$ & The minimum velocity of a gravitating object to form a contrail\\
    $v_z$ & Velocity component in the filament in the $z$-axis direction\\
    $v_\perp$ & Vertical velocity component at the shock wave front\\
    $v_r$ & Velocity in the $r$ direction\\
    $v_\theta$ & Velocity in the $\theta$ direction\\
    $v_\infty$ & Velocity of the gravitating object\\
    $w$ & Filament width\\
    $z_{\rm f}$ & The $z$ coordinate of the filament surface\\
    $\tilde{z}_{\rm f}$ & The $z$ coordinate of the filament surface in units of $r_{\rm HL}$\\
    $\gamma$ & The ratio of specific heats\\
    $\Gamma$ & Heating function\\
    $\Delta v$ & Velocity dispersion inside filament\\
    $\tau$ & Filament formation timescale\\
    $\tau_{\rm d}$ & Filament dispersal timescale\\
    $\mu$ & Mean molecular weight\\
    $\kappa$ & Density enhancement due to orbital compression\\
    $\kappa_{\rm f}$ & Density enhancement due to orbital compression at the filament surface\\
    ${\mathcal K}_{\rm sh}$ & Density enhancement by shock compression at the filament surface\\
    $\Lambda$ & Cooling function\\
    $\rho_\infty$ & Mass density of the ambient medium\\
    \hline
  \end{tabular}
\end{table}

\section{Methods} \label{sec:methods}
\subsection{Orbit of Gas Fragments}
Turbulent structures are ubiquitously observed in the ISM
\citep[e.g.,][]{Miville-Deschenes_etal_2016,Miville-Deschenes_etal_2017}.
To form a linear gaseous structure, it must be formed in a shorter time than eddy turnover time.
In this paper we consider a gravitating object moving at supersonic speeds relative to the ISM.
For simplicity, we consider the gravitating object moving at a constant velocity through an interstellar gas of uniform density. 
We describe the hydrodynamical equations in the comoving frame of the gravitating object. In this frame, we expect that the solution is in a steady state. 
In this section, we neglect the effect of the gas pressure prior to the formation of a shock wave. The effect of the pre-shock pressure will be discussed in \S\ref{Dimless}.

\subsubsection{Characteristic Length Scale}
A characteristic length scale for the accretion onto a gravitating object is called the Hoyle--Lyttleton accretion radius \citep{Hoyle_Lyttleton_1939,Hoyle_Lyttleton_1941,Bondi_Hoyle_1944} and is defined as follows:
\begin{equation}
  \label{eq:r_HL}
  r_{\rm HL}=\frac{2GM}{v_\infty^2},
\end{equation}
where $v_\infty$, $G$, and $M$ are the velocity of the gravitating object, the gravitational constant, and the mass of the gravitating object.

\subsubsection{Orbit's Equation}

\begin{figure}[htbp]
  \begin{center}
   \includegraphics[width=100mm]{ 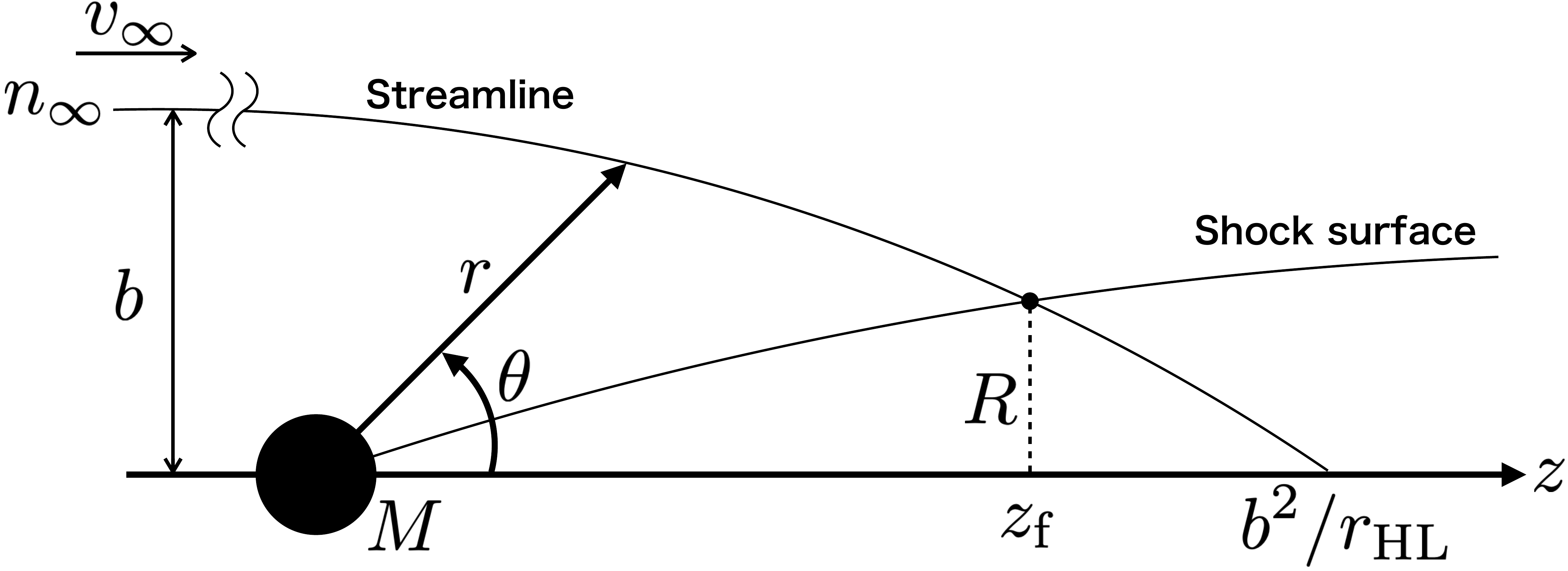}
  \end{center}
   \caption{
   Geometry and sketch.
   \label{fig:geometry}
   }
\end{figure}

First, we derive the gas streamline in the rest frame of the gravitating object.
Since we are considering uniform background density, the solution should be symmetric with respect to the straight path of the gravitating object.
We choose a polar coordinate plane $(r,\theta,\phi)$, and the resultant structure is uniform in the $\phi$ direction.
If we neglect pressure, gas elements with the same impact parameter intersect at a single point and we choose the z-axis as the line connecting the center of the gravitation object to that point.
We consider the geometry shown in Fig.\ref{fig:geometry}.
If the pressure gradient force can be neglected as described above, the streamline is a freefall solution.
The derivation of the orbit of the motion is straightforward \citep[e.g.,][]{GalaDy}. 
Taking the gravitating object as the origin of the coordinates,
the gas equation of motion and the boundary condition are given as

\begin{eqnarray}
  &&\ddot{r}-r\dot{\theta}=-\frac{GM}{r^2},\\
  &&\frac{d}{dt}(r^2\dot{\theta})=0,\\
  &&\lim_{\theta\rightarrow\pi}r\rightarrow\infty,\\
  &&\lim_{\theta\rightarrow\pi}\dot{r}\rightarrow -v_\infty,
\end{eqnarray}
and the streamlines are obtained as follows:

\begin{equation}
  \label{eq:stream}
  r=\frac{b^2v_\infty^2}{GM(1+\cos\theta)+bv_\infty^2\sin\theta},
\end{equation}
where $b$ is the impact parameter of the gas fragments. 
Also, the velocity in the $r$ direction $v_r$ and the $\theta$ direction $v_\theta$ are defined as follows:
\begin{eqnarray}
  v_r^2 &=& \left( 1 + \frac{r_{\rm HL}}{r} - \frac{b^2}{r^2} \right) v_\infty^2,\\
  v_\theta &=& -r\dot{\theta} = +\frac{b}{r}v_\infty,
\end{eqnarray}
where $v_r<0$ for $\theta>\arctan{(2b/r_{\rm HL})}$ and $v_r>0$ for $\theta<\arctan{(2b/r_{\rm HL})}$.

Nondimensionalizing the stream equation using $r_{\rm HL}$ yields
\begin{equation}
  \label{streamline}
  \tilde{r}
  =\frac{\tilde{b}^2}{\frac{1}{2}(1+\cos\theta)+\tilde{b}\sin\theta},
\end{equation}
where $\tilde{r}={r}/{r_{\rm HL}}$ and $\tilde{b}={b}/{r_{\rm HL}}$.
This nondimensional formula is convenient because it does not depend on $M$ or $v_\infty$.

\subsubsection{Compressibility}
By the law of conservation of mass and Eq.(\ref{eq:stream}), the compressibility $\kappa$ due to accretion can be obtained as follows:
\begin{equation}
  \kappa
  =\frac{n}{n_\infty}
  =\frac{b^2}{r\sin\theta(2b-r\sin\theta)},
\end{equation}
where $n$ and $n_\infty$ are the number density at $(r,\theta)$ in the streamlines and the number density of the gas far upstream.
If the gas falls onto the $z$-axis, $\kappa$ will diverge.
In reality, the density does not diverge because the shock wave is generated before the gravitationally accelerated gas falls onto the $z$-axis and the flow is bent to be more in the direction of the $z$-axis.
We define $R$ as the shock surface.
From $R=r\sin\theta$, the compressibility on the surface $\kappa_{\rm f}$ as follows:
\begin{equation}
  \label{kappa}
  \kappa_{\rm f}
  =\frac{b^2}{R(2b-R)}
  =\frac{\tilde{b}^2}{\tilde{R}(2\tilde{b}-\tilde{R})},
\end{equation}
where $\tilde{R}={R}/{r_{\rm HL}}$.

\subsubsection{Accretion Rate per Unit Length of the Filament}
Now we consider the accretion rate per unit length of the filament.
For simplicity, we neglect the shock surface and the filament width.
Thereby, the gas falls onto the $z$-axes and $b^2=r_{\rm HL}\cdot z$ holds (substitute $\theta = 0$ in Eq.(\ref{eq:stream})).
In this approximation, the mass accretion rate per unit length $d\dot{M}$ is as follows:
\begin{equation}
  d\dot{M}=2\pi b\cdot db v_\infty \mu m_{\rm p}n_\infty=\mu m_{\rm p}A dz,
\end{equation}
where
\begin{equation}
\label{eq:A}
  A=\pi r_{\rm HL}v_\infty n_\infty.
\end{equation}
Moreover, $\mu$ and $m_{\rm p}$ are the average molecular weight
and the proton mass; $\mu = 1.4$ is used in this study.
This equation indicates that the accretion rate on the filament is independent of $z$.

\subsection{Shock Surface\label{2.2}}
The location of the shock $R$ is estimated from the law of conservation of mass along the $z$-axis.
We assume a steady state for simplicity, the mass conservation law in the $z$-axis direction is as follows:
\begin{equation}
  \frac{\partial }{\partial z} (\pi R^2 n_{\rm f} v_z) = A,
\end{equation}
where $n_{\rm f}$ and $v_z$ are the macroscopic number density inside the shock surface and the velocity along the $z$-axis.
This yields
\begin{equation}
  \label{conti}
  \pi R^2 n_{\rm f} v_z = A(z-s),
\end{equation}
where $s$ is a stagnation point.
Although it is difficult to determine accurately the location of the stagnation point,
it is known to be $s\approx r_{\rm HL}$\citep[e.g.,][]{Matsuda_etal_2015}.

As we are interested in the length of the filament,
we seek a solution at $z\gg s$.
Since the effect of the gravitating object has almost no effect at $z\gg s$,
$v_z$ is the same as $v_\infty$ and then 
the solution is as follows:
\begin{equation}
  \pi R^2 n_{\rm f} v_\infty = A z.
\end{equation}
We define the density compression ratio by the shock ${\mathcal K}_{\rm sh}$.
In the limit of a weak shock, the compressibility is close to 1, and in the limit of a strong shock wave, it is 4, so $1<{\mathcal K}_{\rm sh}<4$.
Since $n_{\rm f} = {\mathcal K}_{\rm sh}\kappa_{\rm f} n_\infty$ holds, we have
\begin{equation}
  R = \sqrt{\frac{r_{\rm HL}z_{\rm f}}{{\mathcal K}_{\rm sh}\kappa_{\rm f}}},
\end{equation}
where $z_{\rm f}$ is the $z$ value of the intersection of streamline and shock surface.

As in the case of streamlines,
the nondimensionalization is as follows:
\begin{equation}
  \label{haba}
  \tilde{R} = \sqrt{\frac{\tilde{z}_{\rm f}}{{\mathcal K}_{\rm sh}\kappa_{\rm f}}},
\end{equation}
where $\tilde{z}_{\rm f}={z_{\rm f}}/{r_{\rm HL}}$.

Considering that the shock is not so strong far from the gravitating object
and the post-shock region is actually supported by magnetic pressure in the magnetized ISM \citep[e.g.,][]{Inoue_Inutsuka_2009}, 
${\mathcal K}_{\rm sh} \sim 2.0$ can be used in an approximate calculation.

For the sake of later discussion, we derive the vertical velocity component at the shock wave front $v_\perp$ as follows:
\begin{equation}
  \label{vperp}
  v_\perp
  =-\left.
  \frac{v_r\left(-\frac{dR(z)}{dz}\cos\theta +\sin\theta\right) - v_\theta\left(\frac{dR(z)}{dz}\sin\theta+\cos\theta\right)}
  {\sqrt{\left(\frac{dR(z)}{dz}\right)^2+1}}
  \right|_{z=z_{\rm f}}.
\end{equation}

\subsection{Dimensionless Quantities\label{Dimless}}
By Eqs.(\ref{streamline}),(\ref{kappa}), and (\ref{haba}),
the relation between $\tilde{b}$ and $\tilde{z}_{\rm f}$ is as follows:
\begin{equation}
\label{cubic}
  \tilde{z}_{\rm f}^3 + (\tilde{b}^2-1)\tilde{z}_{\rm f}^2-{\mathcal K}_{\rm sh}({\mathcal K}_{\rm sh}+2)\tilde{b}^4 \tilde{z}_{\rm f} + {\mathcal K}_{\rm sh}^2\tilde{b}^6=0.
\end{equation}
The solution is obviously in $0<z_{\rm f}<b^2$.
As mentioned in \S \ref{2.2},
${\mathcal K}_{\rm sh}=2.0$ is used in this study.
From this, we can derive the relationship between dimensionless quantities. 

\begin{figure}[htbp]
    \centering
    \begin{minipage}{0.45\columnwidth}
        \centering
        \includegraphics[width=\columnwidth]{ 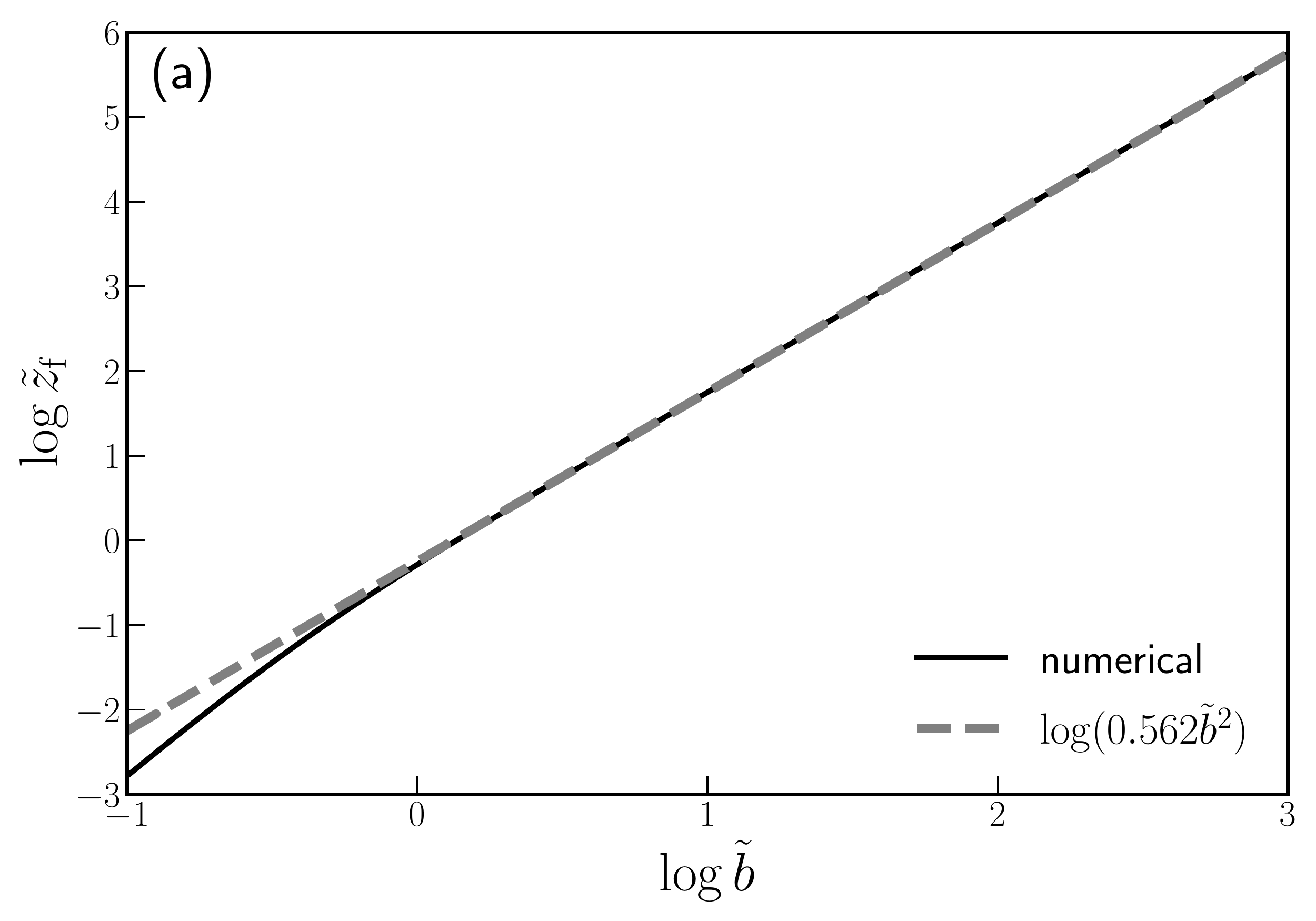}
    \end{minipage}
    \begin{minipage}{0.45\columnwidth}
        \centering
        \includegraphics[width=\columnwidth]{ 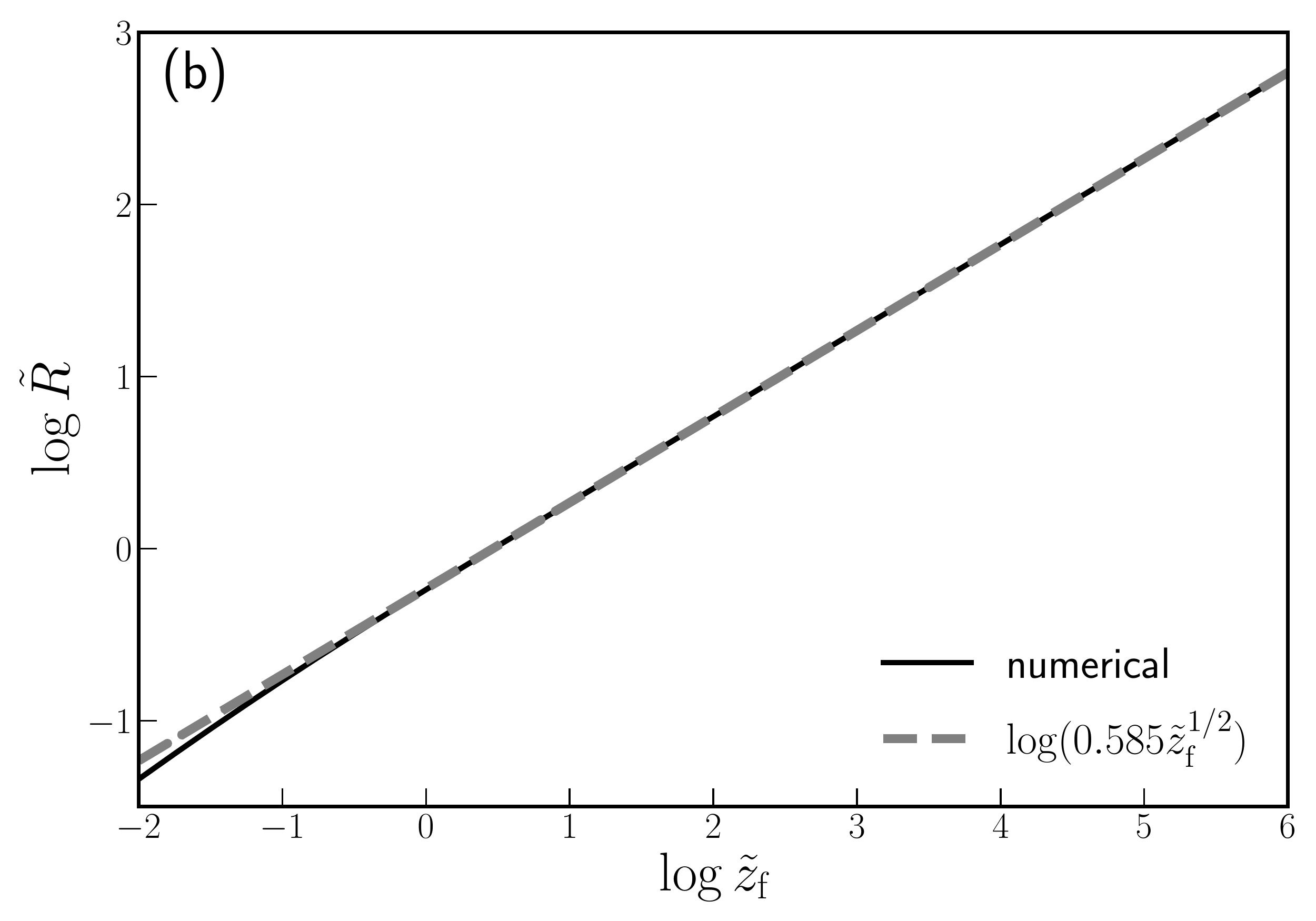}
    \end{minipage}
    \begin{minipage}{0.45\columnwidth}
        \centering
        \includegraphics[width=\columnwidth]{ 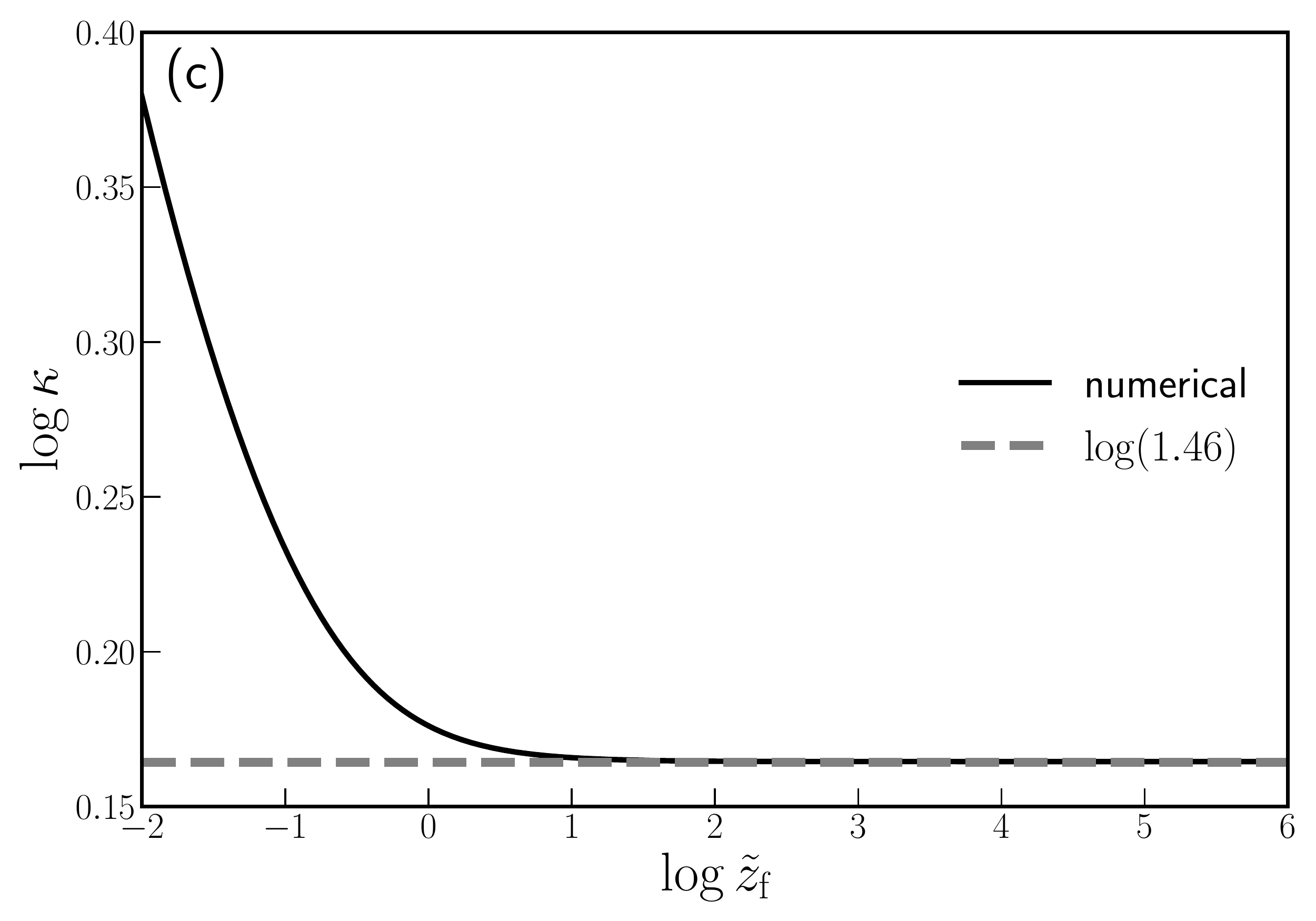}
    \end{minipage}
    \begin{minipage}{0.45\columnwidth}
        \centering
        \includegraphics[width=\columnwidth]{ 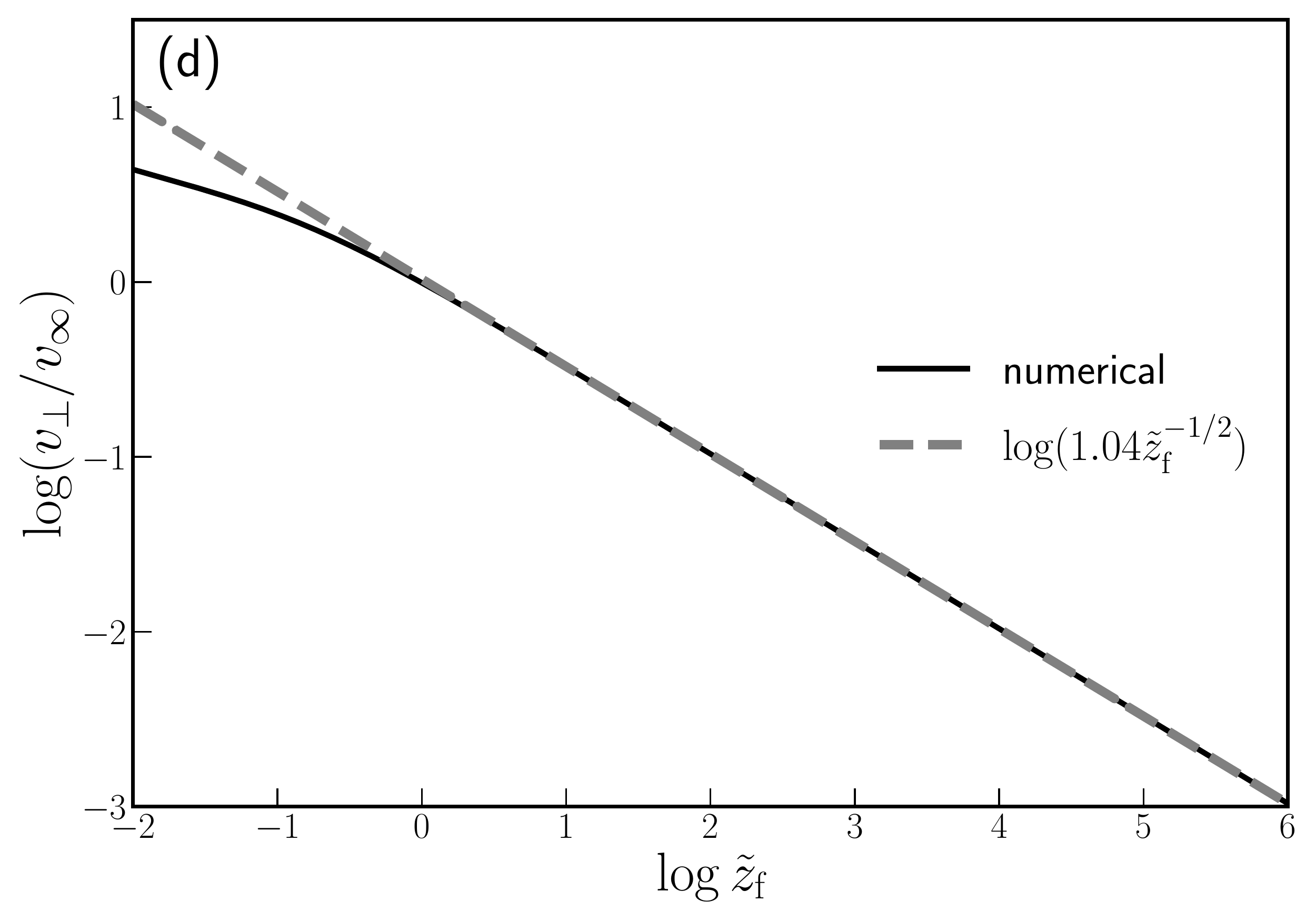}
    \end{minipage}
    \caption
    {
        The relationship between dimensionless quantities in the case of ${\mathcal K}_{\rm sh}=2.0$.
        Since the relation $v_z = v_\infty$ is used, 
        the behavior before the stagnation point $\tilde{z}_{\rm f} \lesssim 1$ is different from the real one.
        The figures show
        (a) the relationship between the dimensionless impact parameter and the position of the filament on the $z$-axis $\tilde{z}_{\rm f}(\tilde{b})$, 
        (b) dimensionless location of the shock surface $\tilde{R}(\tilde{z}_{\rm f})$, 
        (c) compressibility on the shock surface $\kappa_{\rm f}(\tilde{z}_{\rm f})$,
        and (d) dimensionless vertical velocity component at the shock wave front $v_\perp(\tilde{z}_{\rm f})/v_\infty$.
        \label{fig:nondim}
    }
\end{figure}

Now, we are interested in the behavior at the point far from the stagnation point.
In this case, Eq.(\ref{cubic}) is approximated as
\begin{equation}
  \label{z_b}
  \tilde{z}_{\rm f} = \frac{2{\mathcal K}_{\rm sh}}{{\mathcal K}_{\rm sh}+1 + \sqrt{({\mathcal K}_{\rm sh}+1)^2+4{\mathcal K}_{\rm sh}}} \tilde{b}^2.
\end{equation}
From this equation and ${\mathcal K}_{\rm sh}=2.0$,
the values of $\tilde{z}_{\rm f}$, $\tilde{R}$, $\kappa_{\rm f}$ and $v_\perp/v_\infty$ in the region far from the stagnation point are as follows:
\begin{eqnarray}
  \label{eq:1}
  \tilde{z}_{\rm f}
  &\approx& 0.562\tilde{b}^2,\\
  \label{eq:2}
  \tilde{R}
  &\approx&  0.585\tilde{z}_{\rm f}^\frac{1}{2},\\
  \label{eq:3}
  \kappa_{\rm f}
  &\approx& 1.46,\\
  \label{eq:4}
  \frac{v_\perp}{v_\infty}
  &\approx& 1.04\tilde{z}_{\rm f}^{-\frac{1}{2}}.
\end{eqnarray}
From this result, $v_\perp \approx v_\theta(\theta = 0)$ in the region far from the stagnation point (see $v_\theta(\theta = 0)/v_\infty = \tilde{z}^{-1/2}$).
Since $\kappa_{\rm f}$ does not depend on $\tilde{z}_{\rm f}$,
the pressure gradient force can be neglected.
The calculation results are shown in Fig.\ref{fig:nondim}.

\subsection{Length of the Filament}
\subsubsection{Condition of Thermal Condensation}
\begin{figure}[htbp]
\centering
    \begin{minipage}{0.45\columnwidth}
        \centering
        \includegraphics[width=\columnwidth]{ 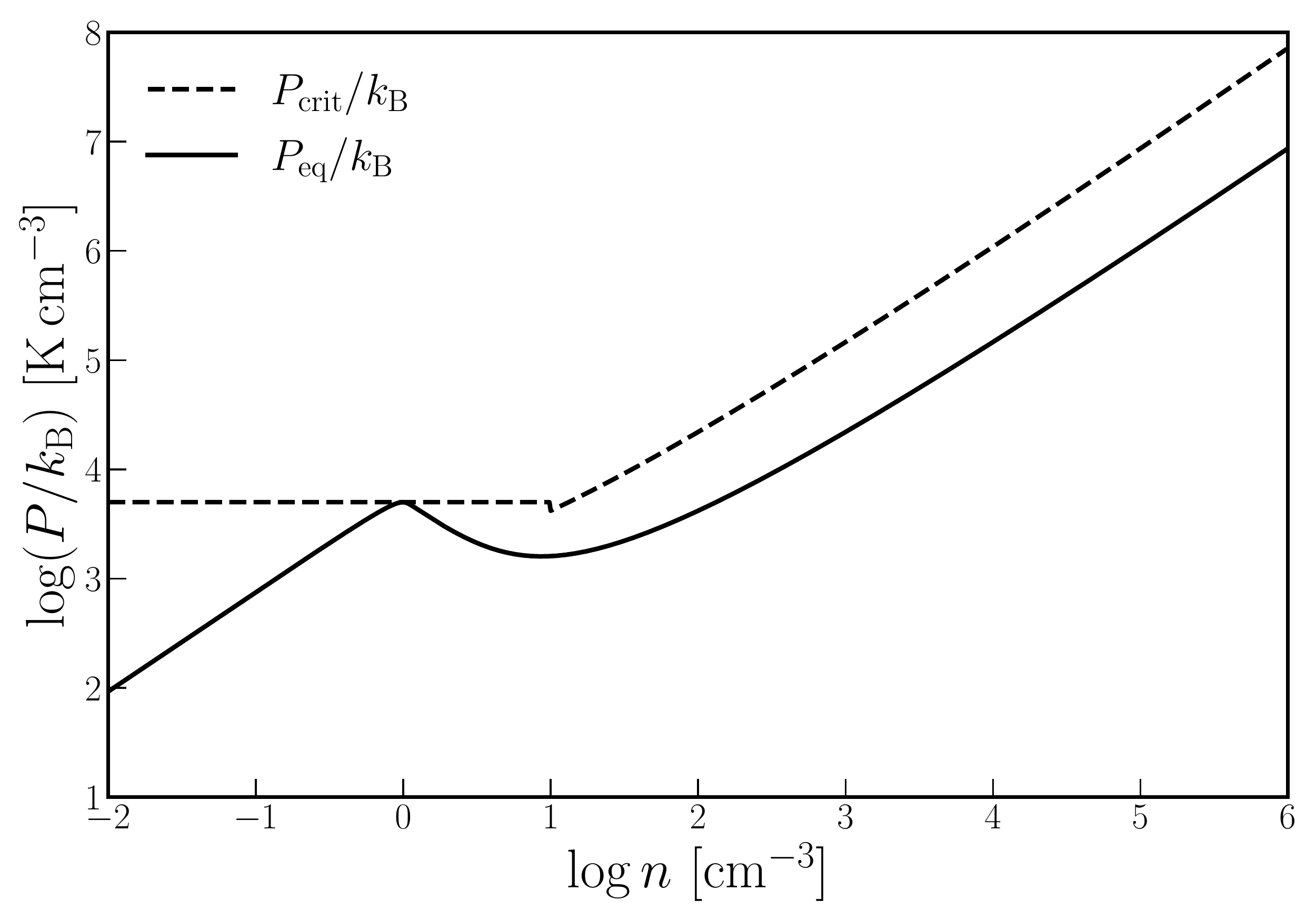}
    \end{minipage}
    \caption
    {
        The critical density $P_{\rm crit}$ and the pressure in thermal equilibrium $P_{\rm eq}$ as a function of number density.
        If the gas pressure increases over $P_{\rm crit}$ by shock, the resultant density grows sufficiently to be observed as a filament after thermal condensation.
        \label{fig:pressure}
    }
\end{figure}

It is known that the evolution of HI gas after the shock compression is almost isobaric and if the post-shock pressure is sufficiently high, the resultant density increases by orders of magnitudes by thermal condensation \citep{Koyama_Inutsuka_2000}.
In reality, the resultant structure is not a one-phase but a multiphase medium where cold high-density material moves turbulently in a warm low-density medium by the nonlinear development of thermal instability \citep{Koyama_Inutsuka_2002,Audit_Hennebelle_2005,Inoue_Inutsuka_2008,Inoue_Inutsuka_2009}.
We expect that such a multiphase medium corresponds to an interstellar contrail.
In this study, we assume that if the gas pressure at the post-shock region exceeds the critical pressure $P_{\rm crit}$, the resultant density increases sufficiently to be observed as a filament after thermal condensation.
We consider that the filament is observed when its density is 10 times denser than the background gas density,
and for the critical pressure
$P_{\rm crit}$, 
we adopt the following:
\begin{equation}
{P_{\rm crit}}=
 \left\{
\begin{array}{ll}
  10^{3.7}k_{\rm B}\ {\rm K\,cm^{-3}}  & \left(n_\infty <10\;{\rm cm^{-3}}\right),\\
  {P_{\rm eq}(n=10 n_\infty)} & \left(n_\infty \geq 10\;{\rm cm^{-3}}\right),
\end{array}
\right.
\end{equation}
where $k_{\rm B}$ is Boltzmann's constant.
$P_{\rm eq}(n)$ is the pressure of HI gas with a number density of $n$ in thermal equilibrium,
which is approximated as
\begin{equation}
  \left\{
 \begin{array}{ll}
  \Gamma=2\times 10^{-26}\ {\rm erg\,s^{-1}},\\
  \frac{\Lambda(T)}{\Gamma}
  =1.0\times 10^7\exp\left( \frac{-118400}{T+1000} \right)
  +1.4\times 10^{-2}\sqrt{T}\exp\left( \frac{-92}{T} \right)\ {\rm cm}^3,\\
  -\Gamma + n\Lambda(T=T_{\rm eq})=0,\\
  P_{\rm eq}(n)=n k_{\rm B} T_{\rm eq},
\end{array}
\right.
\end{equation}
where $\Gamma$, $\Lambda$, $T$, and $T_{\rm eq}$ are a heating function, a cooling function, the HI gas temperature, and the HI gas temperature in thermal equilibrium
\citep{Inoue_Inutsuka_2008}.
The relation between $P_{\rm crit}$ and $P_{\rm eq}$ is shown in Fig.\ref{fig:pressure}.

\subsubsection{Critical Distance for Shock Wave Generation}
We assume that a shock wave is generated if $v_\perp > c_{\rm pre}$;
$c_{\rm pre}$ is the speed of sound in front of the shock wave.
Assuming that the pre-shock state is in radiative equilibrium, we calculate as follows:
\begin{equation}
\label{eq:precs}
  c_{\rm pre}(z_{\rm f})=
  \sqrt{\frac{\gamma k_{\rm B} T_{\rm eq}(n_{\rm pre})}{\mu m_{\rm p}}},
\end{equation}
where $n_{\rm pre} = \kappa_{\rm f}(z_{\rm f})n_\infty$,
and $\gamma$ is the specific heat ratio; $\gamma = 5/3$ is used in this study.
From these,
$L_{\rm sh}$, the $z$ coordinate of the furthest position where a shock wave is generated, is obtained as
\begin{equation}
  v_\perp(z_{\rm f}=L_{\rm sh})=c_{\rm pre}(z_{\rm f}=L_{\rm sh}).
\end{equation}

\subsubsection{Length Limit from Thermal Condensation Condition}
We estimate the largest $z$ coordinate $L_{\rm pt}$ where the gas density reaches high enough to be observed by thermal condensation.
First,
assuming a shock wave is generated,
we derive the distance from this condition $L_{\rm pt}'$.
The pressure inside the filament $P_{\rm f}$ can be obtained using the momentum conservation law.
Because of the angular momentum conservation law, the velocity in the $\theta$ direction is very small far from the gravitating object.
Assuming that the vertical velocity component in the post-shock region (i.e., in the filament) is negligible and noting that $A=\pi r_{\rm HL}v_\infty n_\infty$, the post-shock pressure balances the combination of the ram pressure and pre-shock pressure as
\begin{equation}
\label{eq:Pf}
  P_{\rm f}
  = \mu m_{\rm p} A v_\perp\cdot\frac{1}{2\pi R } + P_{\rm pre}
  =\frac{1}{2}\rho_\infty v_\infty v_\perp \frac{1}{\tilde{R}} + P_{\rm pre},
\end{equation}
where $\rho_\infty = \mu m_{\rm p} n_\infty$ and $P_{\rm pre}$ is the pressure at the shock wave front.

%From $\kappa_{\rm f} \approx 1.46$ in the region far from the stagnation point, the gas state, whether adiabatic or thermal equilibrium, does not significantly affect the results.
Assuming thermal equilibrium conditions,
$P_{\rm pre}$ is as follows:
\begin{equation}
\label{eq:Peq}
  P_{\rm pre}=
  P_{\rm eq}(n_{\rm pre}).
\end{equation}
$L_{\rm pt}'$ is obtained from the following relation:
\begin{equation}
\label{eq:Lpt'}
  P_{\rm f}(z_{\rm f}=L_{\rm pt}')=P_{\rm crit}.
\end{equation}
From these, $L_{\rm pt}$ is obtained from the following relation:
\begin{equation}
  L_{\rm pt} = \min (L_{\rm pt}',L_{\rm sh}).
\end{equation}
Although we adopt a thermal radiative equilibrium value for the pre-shock temperature in Eq.(\ref{eq:precs}) and  pressure in Eq.(\ref{eq:Peq}), the resulting value of $L_{\rm pt}$ does not change much even if we adopt the pressure value of adiabatic compression. This is because $\kappa_{\rm f}\approx 1.46$ is not very large.

\subsubsection{Filament Length}
If gases become thermally unstable by shock compression, they grow into a mixture of CNM and WNM in a turbulent state \citep{Koyama_Inutsuka_2002}.
In this study, we model the resultant shocked gas flowing in the $z$-axis direction as a multiphase filamentary structure that will diffuse out in the region far away from the gravitating object.

We estimate the dispersal timescale $\tau_{\rm d}$.
After the shock compression,
a fraction of the gas becomes a cold dense phase by thermal instability,
but the low-density warm gas tends to be supported by the interstellar magnetic field.
As a result, the average density does not increase much.
Thus, we approximately use $n_{\rm f}$ as the overall macroscopic number density of the gas after the shock compression. 
We assume that the gas expands by turbulence while flowing at $v_\infty$ in the $z$-axis direction
and we define the dispersal time at which the overall decreasing density reaches $n_\infty$.
From the law of conservation of mass, the following equation is obtained:
\begin{equation}
  \pi R^2(L_{\rm pt}) n_{\rm f}=\pi R_{\rm max}^2 n_\infty,
\end{equation}
where $R_{\rm max}$ is the radius at the end of the filament.
From this, $\tau_{\rm d}$ can be estimated as follows:
\begin{equation}
  \tau_{\rm d}=\frac{R_{\rm max} - R(L_{\rm pt} ) }{\Delta v},
\end{equation}
where $\Delta v$ is the dispersion velocity inside the filament,
and we estimate as follows:
\begin{equation}
  \Delta v = \sqrt{\frac{\gamma k_{\rm B} T_{\rm post\mathchar`-shock}}{\mu m_{\rm p}}},
\end{equation}
where $T_{\rm post\mathchar`-shock}$ is the temperature in thermal equilibrium in the post-shock region,
and $T_{\rm post\mathchar`-shock}=T_{\rm eq}(n_{\rm f}(L_{\rm pt}))$.
Using the dispersal time scale we define the filament length in the following:
\begin{equation}
  \label{eq:Length}
  L=L_{\rm pt}+\tau_{\rm d} v_\infty.
\end{equation}

\subsection{Basic Dimensions of the Filament}
We estimate other basic dimensions of the filament such as the width and the total mass.
Since the filament boundary surface depends on the value of $z$,
a representative value is taken as the filament width.
This study estimates the filament width $w$ as follows:
\begin{equation}
  \label{eq:Width}
  w=2R(L_{\rm pt}).
\end{equation}

Next, we approximate the total filament mass $M_{\rm f}$.
We denote $b_{\rm max}=b(L_{\rm pt})$, and
the accretion rate on the filament $\dot{M}_{\rm f}$ is as follows:
\begin{equation}
  \label{eq:TotallMass}
  \dot{M}_{\rm f}=\pi b_{\rm max}^2\rho_\infty v_\infty.
\end{equation}
Since the filament formation timescale $\tau = L/v_\infty$,
$M_{\rm f}$ is obtained as follows:
\begin{equation}
\label{eq:Mf}
  M_{\rm f} = \dot{M}_{\rm f}\tau = \pi b_{\rm max}^2\rho_\infty L.
\end{equation}

\section{Overall Result\label{sec:result}}
\begin{figure}[htbp]
    \centering
    \begin{minipage}{0.45\columnwidth}
        \centering
        \includegraphics[width=\columnwidth]{ 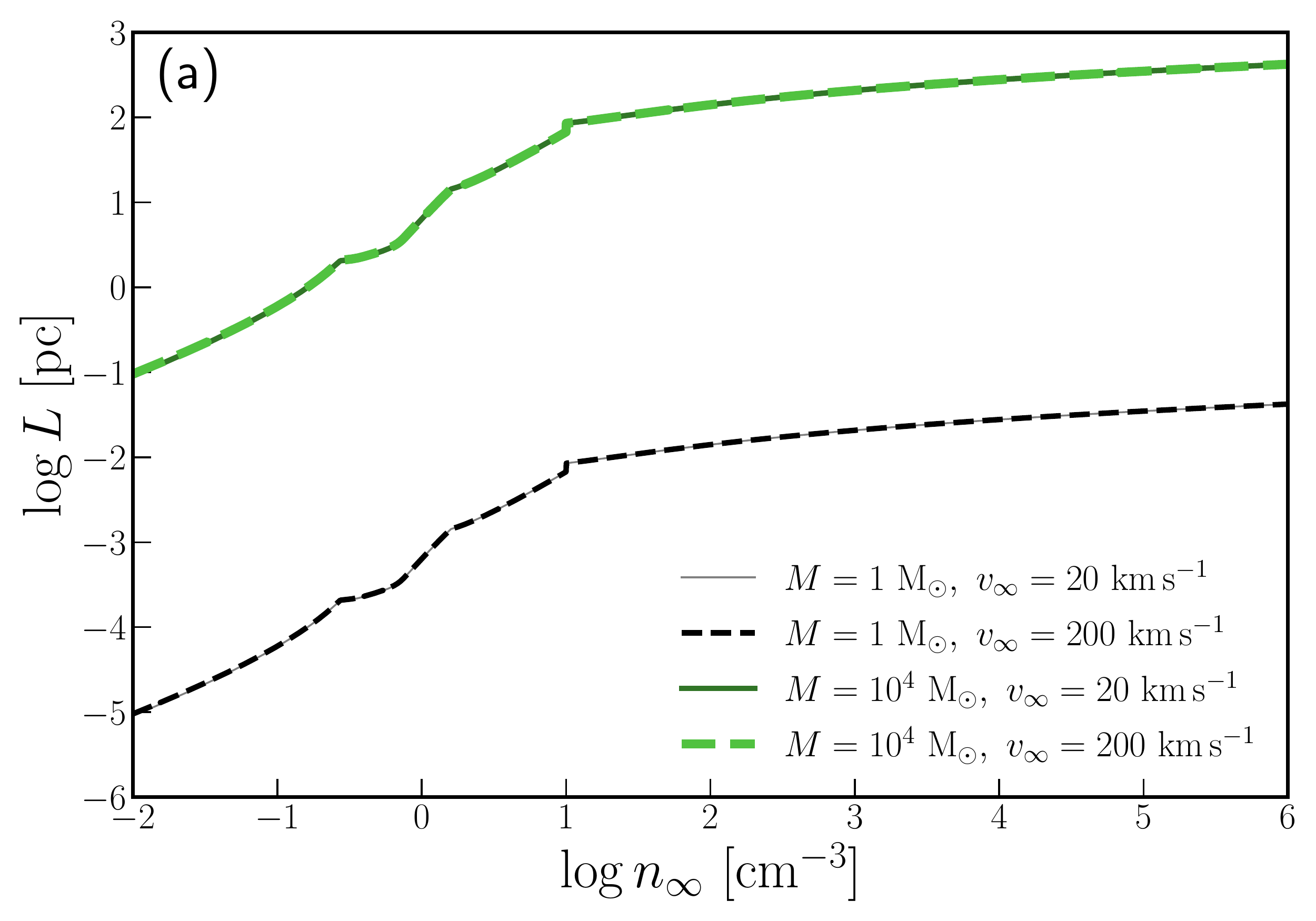}
    \end{minipage}
    \begin{minipage}{0.45\columnwidth}
        \centering
        \includegraphics[width=\columnwidth]{ 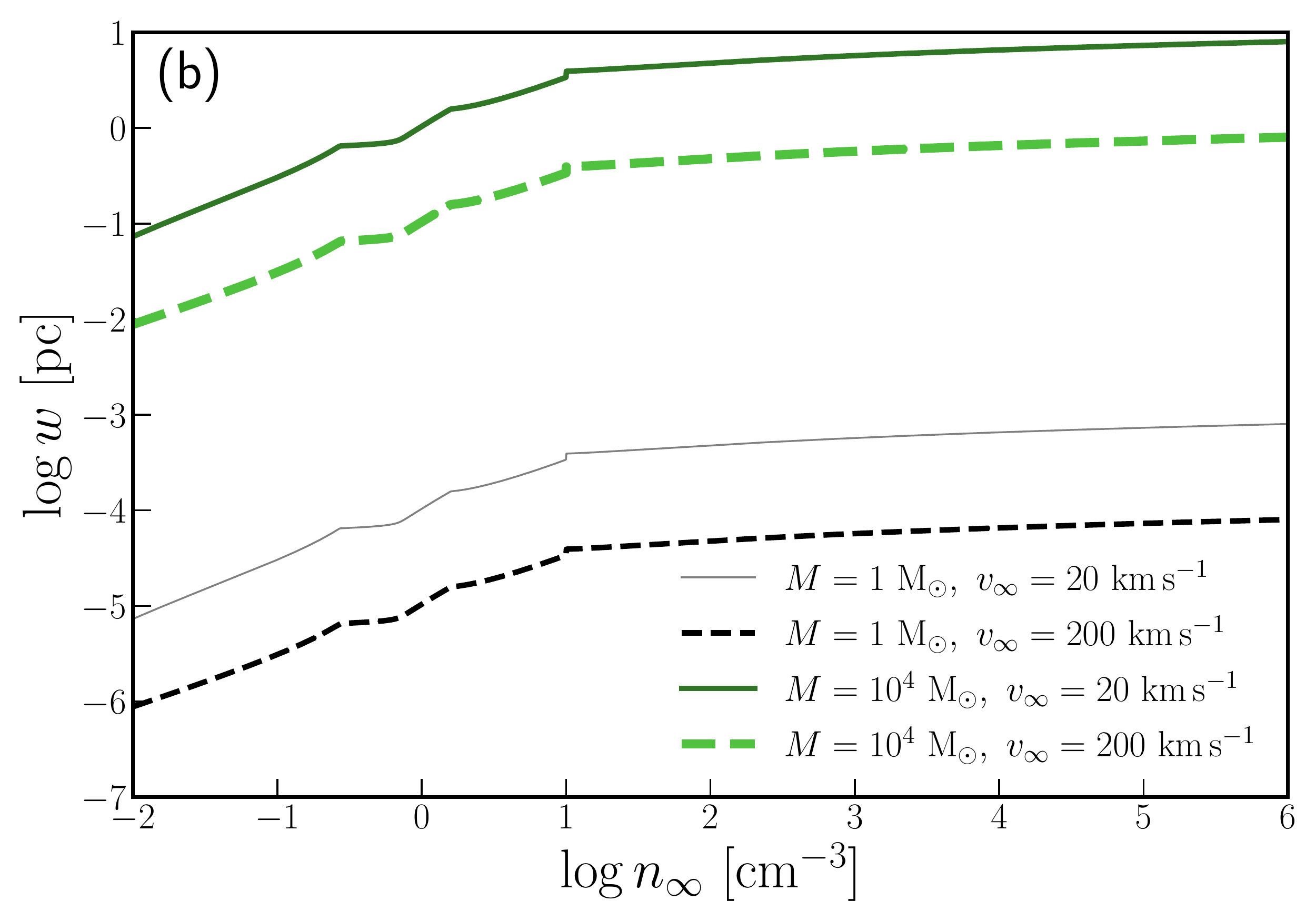}
    \end{minipage}\\

    \begin{minipage}{0.45\columnwidth}
        \centering
        \includegraphics[width=\columnwidth]{ 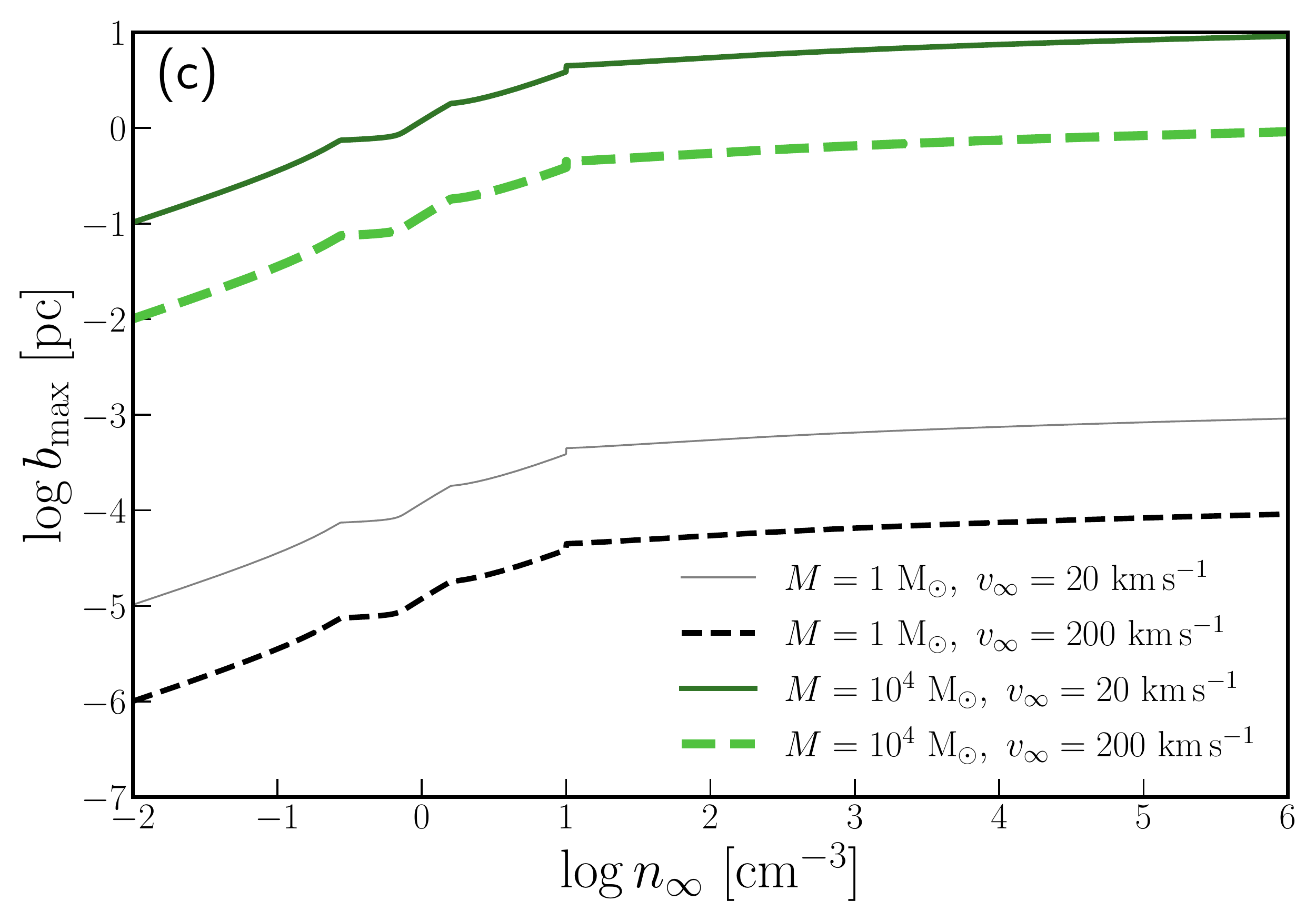}
    \end{minipage}
    \begin{minipage}{0.45\columnwidth}
        \centering
        \includegraphics[width=\columnwidth]{ 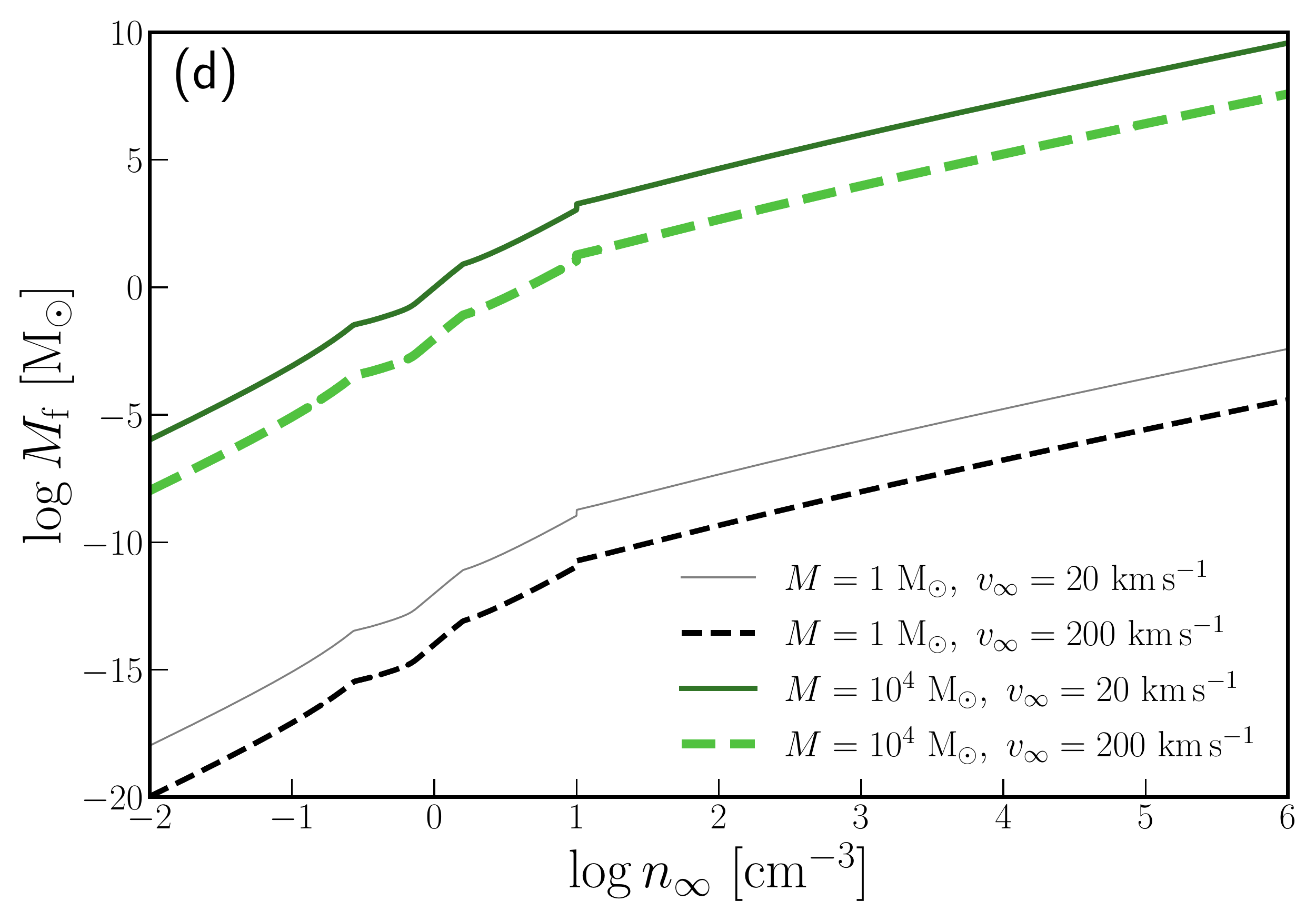}
    \end{minipage}
    \caption
    {
        The basic dimensions of the filament.
        The figures show 
        (a) length $L$, 
        (b) width $w$, 
        (c) $b_{\rm max}$, 
        and (d) total mass $M_{\rm f}$.
        \label{fig:result}
    }
\end{figure}

We calculate the length $L$, the width $w$, $b_{\rm max}$ and the total mass $M_{\rm f}$ from with $M=1,10^4\ {\rm M_\odot}$,
$v_\infty = 20,200\ {\rm km\,s^{-1}}$,
and $10^{-2}\ {\rm cm^{-3}}\leq n_\infty\leq 10^{6}\ {\rm cm^{-3}} $.
We consider a stellar-mass and an intermediate-mass black hole as a gravitating object. 
For the velocity of those objects relative to the ambient ISM, we consider the range from the speed of sound of WNM ($\sim 10\ {\rm km\,s^{-1}}$) and the virial velocity in our galaxy ($\sim$ a few $10^2\ {\rm km\,s^{-1}}$). 
The background gas is assumed to be WNM or CNM. 
A range of densities including WNM and CNM was employed as the density of the background gas \citep[e.g.,][]{Cox_2005,Hennebelle_Inutsuka_2019}.
Fig.\ref{fig:result} shows the result of the calculation.
We find that various filaments are formed depending on the parameters such as mass and velocity of the gravitating object, and background density.
Especially notable is that a compact object more massive than $10^4\ {\rm M_\odot}$ can make a filament whose length is larger than $100\ {\rm pc}$.

This section describes the qualitative behavior.
Quantitative estimates are provided in \S\ref{dis:fil}.

The length is an increasing function of $M$ and $n_\infty$ as shown in Fig.\ref{fig:result}(a) because the momentum flux increases with $M$ and $n_\infty$ as can be seen in Eq.(\ref{eq:A}).
Moreover, the filament length does not depend on $v_\infty$
if $v_\infty$ is sufficiently large ($v_\infty > v_{\rm FC}$; $v_{\rm FC}$ will be discussed in \S \ref{dis:vFC}) because for increasing $v_\infty$,  the effect of  increasing the momentum flux and decreasing the mass accretion rate per unit length $A$ (see Eqs.(\ref{eq:r_HL}) and (\ref{eq:A})) cancel each other.
The filament width $w$ is an increasing function of $M$ and $n_\infty$ and decreasing function of $v_\infty$ as shown in Fig.\ref{fig:result}(b). 
The reason for an increasing function of $M$ and $n_\infty$ is the same as the reason for the length's behavior (see Eq.(\ref{eq:Width})).
The reason for the filament width is a decreasing function of $v_\infty$ where the total mass $M_{\rm f}$ becomes smaller as $v_\infty$ becomes larger as will be explained later and the filament length does not depend on $v_\infty$.

The impact parameter $b_{\rm max}$ that corresponds to $L_{\rm pt}$ is also an increasing function of $M$ and $n_\infty$ and decreasing function of $v_\infty$ as shown in Fig.\ref{fig:result}(c). 
The reason for an increasing function of $M$ and $n_\infty$ is the same as the reason for the length's behavior.
The reason for the decreasing function of $v_\infty$ is derived from $r_{\rm HL}$, which is also a decreasing function of $v_\infty$ and $b^2=r_{\rm HL}\cdot z$ holds (substitute $\theta = 0$ in Eq.(\ref{eq:stream})).

The total filament mass $M_{\rm f}$ is also an increasing function of $M$ and $n_\infty$ and decreasing function of $v_\infty$ as shown in Fig.\ref{fig:result}(d).
The reason for this behavior can be understood from the behavior of $b_{\rm max}$ and $L$ (see Eq.(\ref{eq:Mf})).
The reason for this behavior can also be understood as follows.
If $v_\infty$ is larger, the filament formation timescale $\tau$ is shorter and the accretion rate on the filament $\dot{M}_{\rm f}$ is also small.

\section{Discussion}
\subsection{Approximate Expressions for Various Physical Quantities of the Filaments\label{dis:fil}}
We derive expressions for various physical quantities of filaments.
For simplicity,
we ignore $P_{\rm pre}$ and use the approximation $L\approx 2 L_{\rm pt}'$.
In this approximation, the length $L$ follows from Eqs.(\ref{eq:2}), (\ref{eq:4}) , (\ref{eq:Pf}), and (\ref{eq:Lpt'}) so that
\begin{equation}
\label{eq:length}
  L\approx 2.59\ {\rm pc}
  \left( \frac{M}{10^4\ {\rm M_{\odot}}} \right)
  \left( \frac{n_\infty}{0.5\ {\rm cm^{-3}}} \right)
  \left( \frac{P_{\rm crit}/k_{\rm B}}{10^{3.7}\ {\rm K\,cm^{-3}}} \right)^{-1}.
\end{equation}
Thus, it is shown that $L$ does not depend on $v_\infty$.

Using the approximation $L_{\rm pt} \approx L_{\rm pt}'$, the filament width $w$, $b_{\rm max}$ and total mass $M_{\rm f}$ can be estimated similarly as follows:
\begin{eqnarray}
  w&\approx& 6.17\times 10^{-2}\ {\rm pc}
  \left( \frac{M}{10^4\ {\rm M_{\odot}}} \right)
  \left( \frac{v_\infty}{200\ {\rm km\,s^{-1}}} \right)^{-1}
  \left( \frac{n_\infty}{0.5\ {\rm cm^{-3}}} \right)^{1/2}
  \left( \frac{P_{\rm crit}/k_{\rm B}}{10^{3.7}\ {\rm K\,cm^{-3}}} \right)^{-1/2},\\
  b_{\rm max}&\approx& 7.03\times 10^{-2}\ {\rm pc}
  \left( \frac{M}{10^4\ {\rm M_{\odot}}} \right)
  \left( \frac{v_\infty}{200\ {\rm km\,s^{-1}}} \right)^{-1}
  \left( \frac{n_\infty}{0.5\ {\rm cm^{-3}}} \right)^{1/2}
  \left( \frac{P_{\rm crit}/k_{\rm B}}{10^{3.7}\ {\rm K\,cm^{-3}}} \right)^{-1/2},\\
  M_{\rm f}&\approx& 6.96\times 10^{-4}\ {\rm M_\odot}
  \left( \frac{M}{10^4\ {\rm M_{\odot}}} \right)^3
  \left( \frac{v_\infty}{200\ {\rm km\,s^{-1}}} \right)^{-2}
  \left( \frac{n_\infty}{0.5\ {\rm cm^{-3}}} \right)^3
  \left( \frac{P_{\rm crit}/k_{\rm B}}{10^{3.7}\ {\rm K\,cm^{-3}}} \right)^{-2}.
  \label{eq:filamentmass}
\end{eqnarray}

\subsection{Conditions for Contrail Formation}\label{dis:vFC}
\begin{figure}[htbp]
\centering
    \begin{minipage}{0.45\columnwidth}
        \centering
        \includegraphics[width=\columnwidth]{ 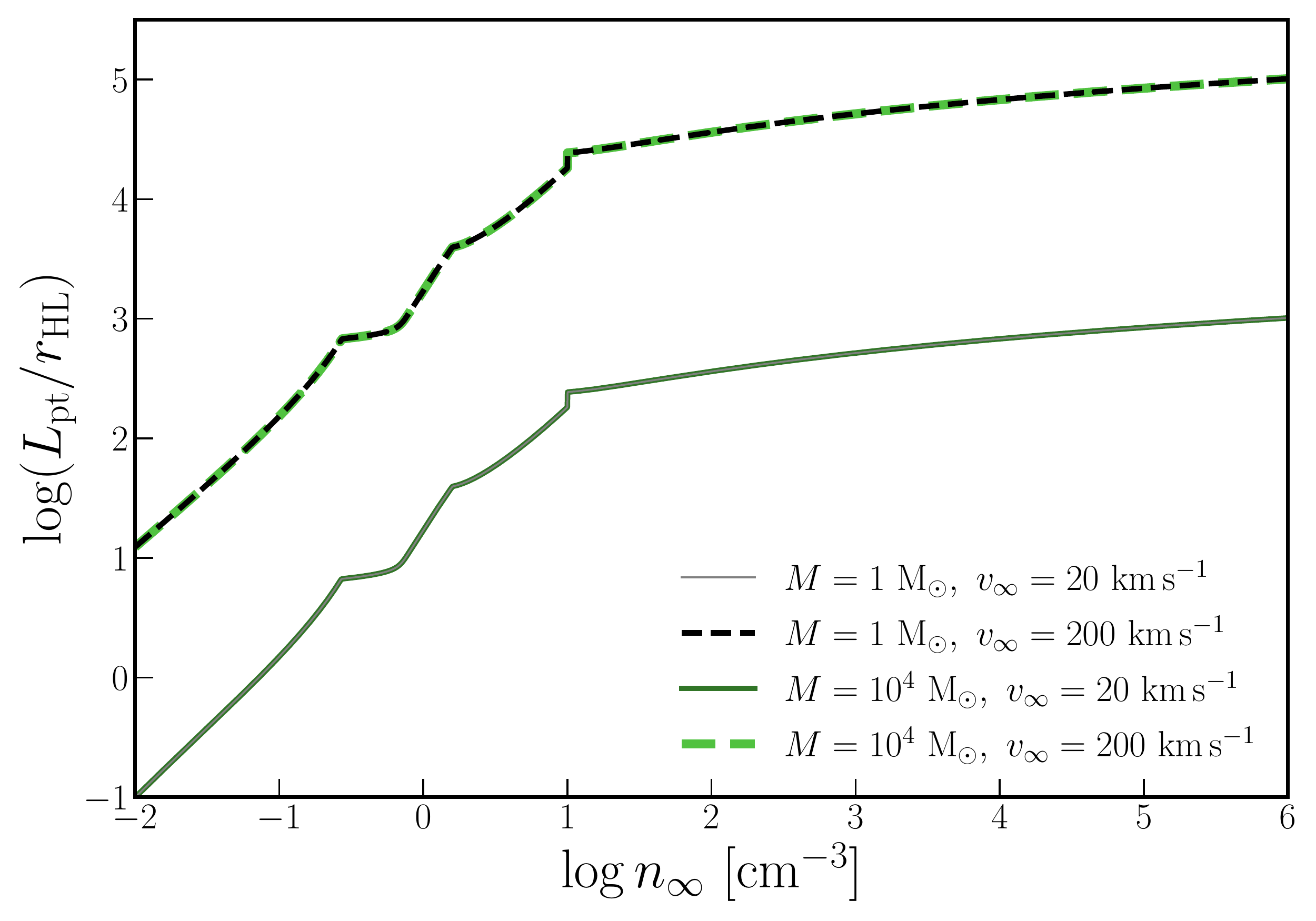}
    \end{minipage}
    \caption
    {
        The ratio of $L_{\rm pt}$ and $r_{\rm HL}$.
        This figure shows that the slower the gravitating object moves, the farther the stagnation point is located from the object.
        If $\log (L_{\rm pt}/r_{\rm HL}) > 0$, we expect that filament should be formed.
        \label{fig:tilde_L}
    }
\end{figure}

First, the radius of the gravitating object should be much larger than $b_{\rm max}$ for creating a contrail.
Second, if a massive compact object moves at a very low velocity, the object only absorbs surrounding gases and obviously does not create a contrail.
In such situations, the stagnation point ($s \approx r_{\rm HL}$) is located far away from the massive object in our framework and the gravitational focusing is negligible. 
If the relative speed $v_\infty$ satisfies $L_{\rm pt} > r_{\rm HL}$, the interstellar contrail is expected to form.
We denote such condition by $v_\infty>v_{\rm FC}$ where $v_{\rm FC}$ is estimated from the approximation $L_{\rm pt} \approx L_{\rm pt}'$ as follows:
\begin{equation}
  v_{\rm FC}\approx 8.15\ {\rm km\,s^{-1}}
  \left( \frac{n_\infty}{0.5\ {\rm cm^{-3}}} \right)^{-1/2}
  \left( \frac{P_{\rm crit}/k_{\rm B}}{10^{3.7}\ {\rm K\,cm^{-3}}} \right)^{1/2}.
\end{equation}
The ratio of $L_{\rm pt}$ and $r_{\rm HL}$ is shown in Fig.\ref{fig:tilde_L}.

\subsection{Filament Formation Site}
The Galactic midplane is supposed to be occupied mostly by WNM except in bubbles due to supernova explosions.
The estimation in \S\ref{dis:fil} corresponds to the case when a massive compact object moves in WNM.
This process is very different from the formation of filamentary molecular clouds in a shock-compressed layer \citep[e.g.,][]{Abe_etal_2021}.

\begin{figure}[htbp]
  \begin{center}
   \includegraphics[width=100mm]{ 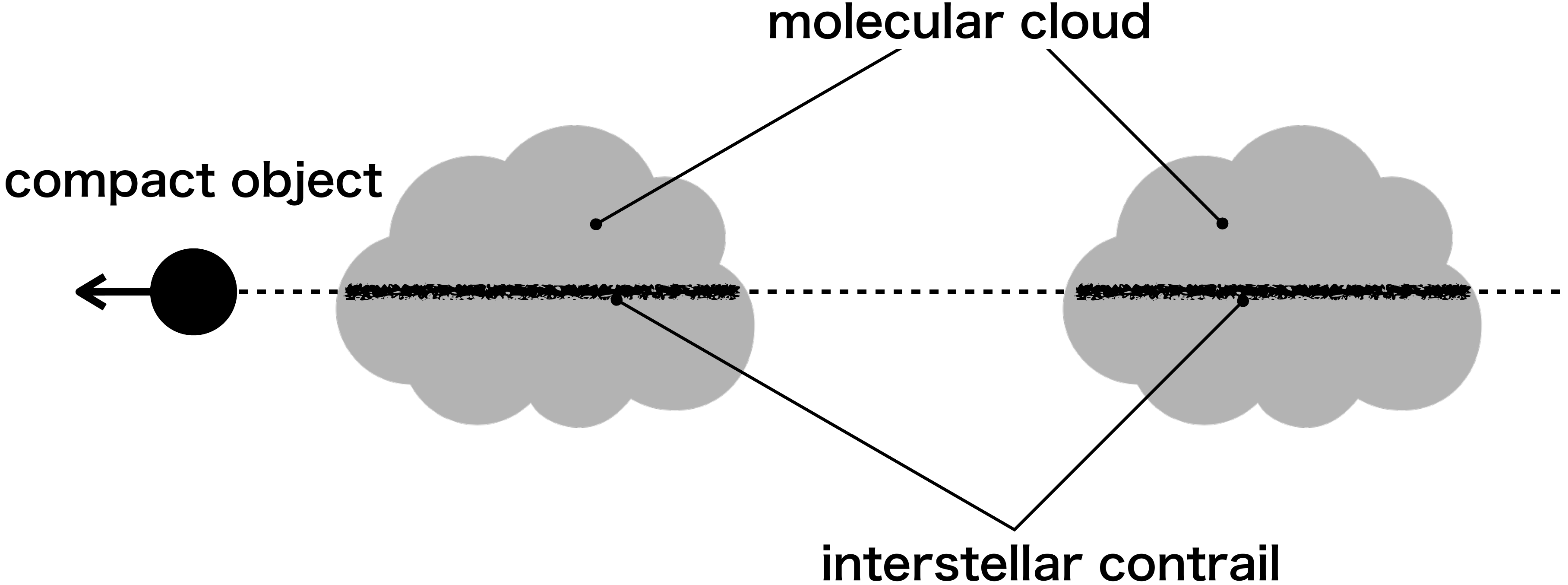}
  \end{center}
   \caption{
   Almost straight but spatially disconnected filaments can be formed by this model.
   The dashed line denotes the pathway of the compact object.
   \label{fig:CNM}
   }
\end{figure}

On the other hand,
a limited volume of the Galactic disk corresponds to CNM (i.e., cold HI clouds and molecular clouds). 
If the CNM size is large enough,
a massive object of $10^4\ {\rm M_\odot}$ may create a long filament as large as $100\ {\rm pc}$, which should be compared to those observed in \citet{Zucker_etal_2018}.
If the CNM size is smaller than $2b_{\rm max}$, 
the filament length is smaller than that estimated in this study.
If the CNM is distributed intermittently on the trajectory of the gravitating source,
the resulting filament might have spatially disconnected (see Fig.\ref{fig:CNM}).

Recent $21\ {\rm cm}$ line observations
\citep[e.g.,][]{HI4PI_2016}
have revealed the existence of numerous linear structures along the interstellar magnetic fields in HI clouds \citep[e.g.,][]{Clark_etal_2014,Jelic_etal_2018,Turic_etal_2021}.
It is difficult to determine the distance to those filaments, which is one of the difficulties in understanding their property.  Although such filamentary structures seem to be ubiquitous on the celestial plane, the cause of those structures is not fully understood. Thus, it is interesting to consider the possibility of creating filamentary structures by multiple gravitating sources passing through the shell of the Local Bubble. 
The typical length of observed HI filaments is roughly estimated as $1\text{--}10\ {\rm pc}$ if they are located on the surface of the Local Bubble whose radius is about $50\ {\rm pc}$
\citep[e.g.,][]{Zucker_etal_2022}.
With Eqs. (\ref{eq:length}) -- (\ref{eq:filamentmass}) we can estimate the condition for a gravitating object to form these filaments.
For example, we consider a gravitating object passing through the Local Bubble's shell. If the number density of ambient gas is $20\ {\rm cm^{-3}}$, gravitating objects of $10$, $10^2$, and $10^3\ {\rm M_\odot}$ create filaments with lengths of approximately $0.078$, $0.78$, and $7.8\ {\rm pc}$, respectively.
Whereas the number density is $200\ {\rm cm^{-3}}$, gravitating objects of $10$, $10^2$, and $10^3\ {\rm M_\odot}$ create approximately $0.14$, $1.4$, and $14\ {\rm pc}$, respectively.
Note that $P_{\rm crit}$ is not a simple linear function of $n_\infty$ so the filament length (Eq. (\ref{eq:length})) has a nonlinear dependence on $n_\infty$.
We should note, however, that the thermal condensation is hindered by the interstellar magnetic field \citep[e.g.,][]{Inoue_Inutsuka_2008,Inoue_Inutsuka_2009}, and the length of the interstellar contrail becomes shorter than our model if the strength of the magnetic field is significantly large.
It is difficult to treat the effects of magnetic fields in the present analytical framework and we will study them by numerical simulation in the future.

\subsection{Accretion to Gravitating Source}
Accretion onto a moving gravitating source is called Bondi--Hoyle--Littleton accretion
\citep{Hoyle_Lyttleton_1939,Hoyle_Lyttleton_1941,Bondi_Hoyle_1944},
and the mass accretion rate to the gravitating source can be estimated by
\begin{eqnarray}
  \dot{M}&=&\pi r_{\rm HL}^2\rho_\infty v_\infty\\
  &\approx& 5.13\times 10^{-11}\ {\rm M_\odot\,yr^{-1}}
  \left( \frac{M}{10^4\ {\rm M_{\odot}}} \right)^2
  \left( \frac{v_\infty}{200\ {\rm km\,s^{-1}}} \right)^{-3}
  \left( \frac{n_\infty}{0.5\ {\rm cm^{-3}}} \right).
\end{eqnarray}
This value seems to be too small to be observed.
A detailed discussion on observability is beyond the scope of this paper \citep[see, e.g.,][]{Fukue_Ioroi_1999,Ogata_etal_2021}.

\section{Conclusions}
We propose a new mechanism for creating an interstellar contrail
formed by the thermal condensation of HI gas or molecular gas in the trajectory of a fast-moving gravitating source in the ISM, and estimate the characteristics of the interstellar contrail analytically.
The results show that the filament length is independent of the relative velocity of the gas and the gravitating object if the formation conditions are satisfied.
It is also found that filaments with a variety of lengths, widths, and masses can be formed depending on the mass of the gravitating source, relative velocity, and the density of ambient gas.
The resulting filament is expected to be in thermal equilibrium and the average number density depends on $P_{\rm crit}$.  Roughly speaking, 
if the gravitating object passes through in WNM, the number density of the filament is on the order of 100 ${\rm cm^{-3}}$ and if the gravitating object passes through in the dense region such as CNM or molecular cloud, the number density of the filament is about 1000 ${\rm cm^{-3}}$ or more.
For example, when a $10^2\ {\rm M_\odot}$ object passes through the dense region ($n_\infty \sim 200\ {\rm cm^{-3}}$) in the shell of the Local Bubble, it creates a filament whose length is about $1.4\ {\rm pc}$.
In particular, if a compact object more massive than $10^4\ {\rm M_\odot}$ such as an intermediate-mass black hole goes through a large region of mostly cold atomic or molecular gas, it can form a filament whose length is larger than a hundred parsecs.
If we observationally identify such phenomena in multiple regions, we will be able to estimate the frequency of intermediate-mass black holes that are not luminous enough to be visible.
This line of work is expected to provide an important step toward understanding the formation process of massive black holes.

\begin{acknowledgements}
This work was financially supported by JST SPRING, Grant Number JPMJSP2125. The
author (K.K.) would like to take this opportunity to thank the “Interdisciplinary Frontier Next-Generation Researcher Program of the Tokai Higher Education and Research System.
The author (S.I.) is supported by JSPS Grants-in-Aid for Scientific Research Nos. 16H02160, 18H05436, and 18H05437.
We thank the anonymous referee for constructive comments. We also thank J. Shimoda for the useful discussion.
\end{acknowledgements}

\clearpage

\bibliography{InterstellarContrails}{}
\bibliographystyle{aasjournal}

\end{document}